\documentclass[10pt, colorlinks,linkcolor=blue,square, citecolor=blue]{elsarticle}

\journal{} 

\makeatletter
\def\ps@pprintTitle{%
  \let\@oddhead\@empty
  \let\@evenhead\@empty
  \let\@oddfoot\@empty
  \let\@evenfoot\@empty}
\makeatother

\usepackage[utf8]{inputenc}

\usepackage{amssymb}
\usepackage{amsmath,mathrsfs}
\usepackage{empheq}

\usepackage[sc]{mathpazo}
\usepackage{bm}

\usepackage{color}

\usepackage{soul}
\soulregister\cite7
\soulregister\ref7 
\soulregister\citep7 
\soulregister\citet7 
\soulregister\pageref7 

\usepackage{extarrows}

\usepackage{verbatim}

\usepackage{siunitx}

\usepackage{stfloats}

\usepackage{bm}

\usepackage{graphicx}

\usepackage{caption}
\usepackage{subcaption}

\usepackage{multicol}

\usepackage{multirow}

\usepackage{mhchem}

\usepackage[toc,page]{appendix}

\usepackage{bm}

\usepackage{float}

\usepackage{enumerate}

\usepackage{threeparttable}

\usepackage{sectsty}
\usepackage[toc,page]{appendix}
\usepackage[colorlinks,linkcolor=blue,anchorcolor=green,citecolor=blue]{hyperref}

\renewcommand{\eqref}[1]{Equation~$($\ref{#1}$)$}

\usepackage{cuted}
\usepackage{flushend}
\usepackage{booktabs} 

\renewcommand{\eqref}[1]{Eq.~$($\ref{#1}$)$}

\begin{document}
	
\begin{frontmatter}
    \title{Phase-field investigation of non-isothermal solidification coupled with melt flow dynamics}
\author[els]{Timileyin David Oyedeji\corref{cor1}}
\ead{oyedeji@mma.tu-darmstadt.de}

\author[elm]{Aaron Brunk}
\author[elp]{Yangyiwei Yang}
\author[rtt]{Herbert Egger}
\author[els]{Holger Marschall}
\author[elp]{Bai-Xiang Xu\corref{cor1}}
\ead{xu@mfm.tu-darmstadt.de}

\cortext[cor1]{Corresponding author}
			
\address[els]{Computational Multiphase Group, Technische Universit\"at Darmstadt, Darmstadt 64287, Germany}

\address[elm]{Institute of Mathematics, Johannes Gutenberg-University Mainz, Germany}

\address[elp]{Mechanics of Functional Materials Division, Institute of Materials Science, Technische Universit\"at Darmstadt, Darmstadt 64287, Germany}

\address[rtt]{Johann Radon Institute for Computational and Applied Mathematics and Institute for Computational Mathematics, Johannes-Kepler University Linz, Austria}

		
\begin{abstract}
Solidification, coupled with melt flow, plays a critical role in determining the microstructure and properties of materials in several manufacturing processes. Phase-field models coupled with the Navier-Stokes equations are widely used to model and simulate these dynamics. However, most existing models neglect essential thermodynamic couplings, particularly the capillary (Korteweg) stress in the momentum equation. This stress, which arises from the coupling between the phase field and the melt flow, accounts for thermal capillary effects during non-isothermal solidification. Neglecting it leads to models inconsistent with non-equilibrium thermodynamics and incapable of capturing capillarity-driven melt flow. In this work, we present a thermodynamically consistent, non-isothermal phase-field model for solidification coupled with melt flow, incorporating cross-coupling terms and explicitly including the Korteweg stress in the momentum equation. Model validation is performed for solidification-only cases, followed by simulations of dendritic growth under melt flow. The results show that thermal capillary effects induce flow near the interface, influencing dendrite tip velocity and morphology. Simulations under forced convection further demonstrate asymmetric dendrite growth due to the imposed flow field. Additionally, we numerically demonstrate the influence of viscosity interpolation schemes on enforcing the no-slip boundary condition in phase-field models with melt flow.
\end{abstract}
		
		
\end{frontmatter}

\section{Introduction}

Solidification is a fundamental phenomenon in various manufacturing techniques, such as casting and additive manufacturing, where it strongly influences microstructural evolution and defect formation in manufactured components. These microstructural features, particularly dendritic growth, directly affect the final properties of the material \cite{takaki2014phase, dantzig2016solidification, rojas2022phase, qiu2024phase}. Solidification involves a complex interaction of multiple physical phenomena such as phase transition, heat and solute transfer, and melt flow, all of which influence microstructure formation to varying degrees. Among these, melt flow plays a critical role as it redistributes the diffusion fields in the liquid phase, thereby affecting growth kinetics and morphology of the microstructure \cite{dantzig2016solidification, sun2019anisotropic, sakane2021phase}. Understanding the coupled dynamics of solidification and melt flow is thus essential for both scientific investigation and process optimization. 

Experimental investigation of solidification and melt flow dynamics remains challenging due to the difficulty of directly observing melt flow and its effects during solidification, especially under high-temperature, transient, and multiscale conditions \cite{qiu2024phase, sakane2021phase, ludwig2005situ}. Moreover, in situ observation techniques are often restricted to thin samples, limiting their applicability \cite{takaki2014phase, rojas2022phase}. Numerical modeling thus serves as a powerful alternative, enabling detailed investigation of the underlying physical phenomena and their interactions \cite{wu2024unified}. In this regard, the phase-field method has emerged as a widely used and versatile approach for simulating solidification phenomena. By representing microstructural evolution through variables typically called order parameters, the phase-field method avoids explicit interface tracking while effectively capturing essential interfacial dynamics. In addition, phenomena such as interface anisotropy and the kinetic effects of interfacial motion can be directly incorporated into the governing equations, allowing for dendritic growth modeling and simulations \cite{kobayashi1993modeling, karma1998quantitative, rojas2015phase}. 

Phase-field modeling of solidification coupled with melt flow typically involves coupling the governing equations for the phase-field order parameters, thermal and/or solutal diffusion, and the incompressible Navier–Stokes equations. This integrated framework captures the complex interaction between phase transition, heat and solute transfer, and melt flow, and has been extensively used to investigate the dynamics of solidification and its interaction with melt flow \cite{diepers1999simulation,nestler2000phase, beckermann1999modeling, anderson2000phase, lu2005three, tong2001phase, yuan2010dendritic, gong2019quantitative}. To enforce the no-slip boundary condition on the solid side of the diffusive interface, various approaches have been developed \cite{subhedar2020thin}. One common method is the variable viscosity model, where viscosity increases smoothly across the interface and diverges in the solid phase \cite{tonhardt1998phase, nestler2000phase, anderson2000phase}. Another approach employs a dissipative interfacial force to impose the boundary condition \cite{beckermann1999modeling, subhedar2015modeling}. 

Moreover, the thermodynamically consistent formulation of the phase-field–Navier–Stokes system, based on non-equilibrium thermodynamics, introduces additional cross-coupling terms in the governing equations \cite{anderson2000phase, yang2020non}. These include couplings between phase-field order parameters and diffusion fields commonly associated with antitrapping effects \cite{brener2012kinetic, boussinot2013interface}, and the capillary stress contribution in the momentum equation, known as the Korteweg stress tensor \cite{anderson2000phase, yang2020non, anderson1998diffuse}. The Korteweg stress originates from the coupling between the phase-field and the melt velocity field, with direct dependence on the interface energy. It not only ensures thermodynamic consistency but also inherently accounts for capillarity-driven fluid flow phenomena \cite{anderson2000phase, onuki2007dynamic}. Notably, under non-isothermal conditions, it captures thermal capillary effects induced by temperature gradients at the interface \cite{li2019phase}. However, the Korteweg stress is often neglected in the modeling and simulation of solidification coupled with melt flow dynamics. This disregards both the thermodynamic consistency of the coupled model and the influence of capillary-driven effects on the evolving interface morphology.

In this work, we present a thermodynamic-consistent non-isothermal phase-field model for solidification
coupled with melt flow dynamics. The explicit derivations inherently contains the cross-coupling effects between the phase-field order parameters as well as the Korteweg stress in the momentum equation. The no-slip boundary condition at the solid–liquid interface is enforced using the variable viscosity model \cite{tonhardt1998phase, nestler2000phase, anderson2000phase}. We first validate the model through numerical simulations of solidification-only cases, including planar interface evolution and dendritic growth, comparing the results to theoretical predictions. Subsequently, we investigate dendritic growth coupled with melt flow, focusing on the flow induced by thermal capillary effects and its influence on dendritic morphology and growth.  Additionally, we examine dendritic growth under forced convection and also numerically investigate the impact of different viscosity interpolation schemes on the enforcement of the no-slip boundary condition. The paper is structured as follows. Section~\ref{model_deriv} presents the thermodynamically consistent model formulation, including the entropy and free energy functionals, and the governing equations for phase transition, heat transfer, and melt flow. Section~\ref{model_result} discusses the numerical results, including model verification and simulations of dendritic growth under various flow conditions. Finally, conclusions are presented in Section~\ref{model_conc}.

\section{Phase-field model} \label{model_deriv}

\subsection{Entropy and Free-energy Functionals}
For the phase-field modeling of non-isothermal solidification coupled with melt flow dynamics and anisotropic interface energy, a phase-field variable $\phi$ is introduced to represent the solid and liquid phases: $\phi = 1$ in the solid phase and $\phi = 0$ in the liquid phase. Considering a finite domain $\Omega$, the entropy functional $S$ for the domain can be defined as
\begin{equation}
    S(e, \phi) = \int_\Omega \left[s(e,\phi) - \frac{1}{2}\kappa_\phi\Gamma^2(\nabla\phi)  \right] \mathrm{d}\Omega, \label{entropy}
\end{equation}
where $s$ is the local entropy density, $e$ is the internal energy density, and $\kappa_\phi$ is the gradient energy coefficient. The function $\Gamma:\mathbb{R}^d\to\mathbb{R}$ accounts for the anisotropic solid-liquid interface energy. If an isotropic interface energy is desired, $\Gamma$ simplifies to the form $\Gamma=|\nabla\phi|$ \cite{taylor1998diffuse, anderson2000phase}.

\noindent Moreover, the internal energy density, $e(T,\phi)$ is formulated as 
\begin{equation}
  e(T,\phi) = e_{s}(T)h(\phi) + e_{l}(T)(1-h(\phi)) = \mathcal{L}(T)h(\phi) + e_{l}, \label{ine}  
\end{equation}
where $e_{s}$ and $e_{l}$ are the internal energy densities of the solid and liquid phases respectively, $\mathcal{L} = e_{s} - e_{l}$ is the latent heat of the liquid-solid transition, and $T$ is temperature. $h(\phi) = \phi^3(6\phi^2 - 15\phi +10)$ is an interpolation function that satisfies; $h(\phi = 0) = 0$ and $h(\phi = 1) = 1$. Following Legendre transformation, the free energy functional $F$ can be obtained as
\begin{equation}
    F(T, \phi) = \int_\Omega \left[f(T,\phi) + \frac{1}{2}T\kappa_\phi\Gamma^2(\nabla\phi)  \right] \mathrm{d}\Omega, \label{fed}
\end{equation}
with the local free energy $f(T,\phi)$ defined as
\begin{equation}
  f(T,\phi) = T \int e (T,\phi) \mathrm{d}\left( \frac{1}{T} \right). \label{lf}  
\end{equation}

\noindent Moreover, based on the procedure in Refs.~\cite{penrose1990thermodynamically,andersson2002phase}, $f(T,\phi)$ can be further defined in terms of a heat term $f_\mathrm{ht}(T)$ and a double well potential $f_\mathrm{dw}(\phi)$,
\begin{equation}
    f(T,\phi) = T (f_\mathrm{ht}(T,\phi) + f_\mathrm{dw}(\phi)),
\end{equation}
with
\begin{equation}
    f_\mathrm{ht}(T,\phi) = - \left[ c_p \ln \frac{T}{T_{M}} +  (h(\phi)-1)\mathcal{L}\left(\frac{1}{T_M}-\frac{1}{T}\right) \right],
\end{equation}
and 
\begin{equation}
    f_\mathrm{dw}(\phi) = H \phi^2 (1-\phi)^2,
\end{equation}
where $T_M$ is the melting temperature, $c_{p}$ is the specific heat capacity, and $H$ is the height of the double well potential. Also, the latent heat $\mathcal{L}$ is simply taken as a constant $\mathcal{L}(T) = \mathcal{L}(T_M) = \mathcal{L}$, implying equal specific heat capacities in the liquid and solid phases; $c_{p} = c_{s} = c_{l}$.  It can be noted that the local free energy is defined such that the driving force for interfacial movement is proportional to the undercooling at the interface. 


\subsection{Kinetic Equations}
Taking into account melt flow dynamics via introducing the velocity field $\mathbf{u}$, the total energy density of the system, $e_{\text{tot}} (T,\phi, \mathbf{u})$ can be defined as the sum of the internal energy density $e (T,\phi)$ and kinetic energy density $\mathrm{k}_e(\mathbf{u})$ \cite{Yang2019, jacqmin1999calculation, yang2020non}:
\begin{equation}
    e_{\text{tot}}(T,\phi, \mathbf{u}) = e (T,\phi) + \mathrm{k}_e(\mathbf{u}) = e (T,\phi) + \frac{1}{2}\varrho|\mathbf{u}|^2,
\end{equation}
where $\mathbf{u}$ is the melt flow velocity, and $\varrho$ is the density taken to be identical and constant in both the liquid and solid phases. Therefore, the melt flow can be described using the incompressible Navier-Stokes equations:
\begin{align}
    \varrho\frac{\mathrm{D} \mathbf{u}}{\mathrm{D} t}&= \nabla \cdot \pmb{\sigma}  + \mathbf{b}, \qquad\qquad \nabla \cdot \mathbf{u} = 0, \label{ns}
\end{align}
where $\pmb{\sigma}$ is the Cauchy stress tensor, and $\mathbf{b}$ is the body force. The term ${\mathrm{D} (\cdot)}/{\mathrm{D} t} = (\cdot) + (\mathbf{u} \cdot \nabla) (\cdot)$ denotes the material derivative. Following Ref.~\cite{yang2020non}, the local energy conservation within the domain can be formulated as 
\begin{equation}
   \frac{\mathrm{D} e}{\mathrm{D} t} = \pmb{\sigma} : \nabla\mathbf{u} - \nabla\cdot\mathbf{J}_{e} \label{lec},
\end{equation}
where $\mathbf{J}_{e}$ is the internal energy flux. For the phase-field variable $\phi$, we propose an Allen-Cahn type equation,
\begin{equation}
    \frac{\mathrm{D} \phi}{\mathrm{D} t} = J_\phi, \label{phiev}
\end{equation}
where $J_{\phi}$ is the driving force yet to be determined.
To ensure non-negative entropy production, we employ Onsager relations, which establish linear relationships between thermodynamic fluxes, non-conserved time evolution equations, and their corresponding thermodynamic driving forces. This yields \cite{yang2020non, oyedeji2023variational}:
\begin{align}
    \mathbf{J}_{e} = \mathbf{L}_{\phi e}\frac{\delta S}{\delta \phi} + \mathbf{L}_{ee}\nabla \frac{\delta S}{\delta e}, \label{efe}
\end{align}
\begin{align}
    J_\phi= \mathbf{L}_{\phi \phi}\frac{\delta S}{\delta \phi} + \mathbf{L}_{\phi e}\cdot\nabla \frac{\delta S}{\delta e}. \label{efx}
\end{align}
The Onsager mobility matrix $\mathbf{L}$ is notably symmetric, and its components can depend on $\phi$, $T$, and their spatial derivatives. Dimensionally, $\mathbf{L}_{\phi\phi}$ is a scalar function, $\mathbf{L}_{\phi e}$ is a $d$ dimensional vector and $\mathbf{L}_{e e}$ is a $d\times d$ matrix.
Furthermore, the constitutive equation for the Cauchy stress tensor $\pmb{\sigma}$ is given by
\begin{equation}
    \pmb{\sigma} = -p\mathbf{I} + \nu \nabla_{\text{s}}\mathbf{u} - \pmb{\sigma}_\mathrm{K},\label{str1}
\end{equation}
with
\begin{equation}
    \pmb{\sigma}_\mathrm{K} = T\kappa_\phi\Gamma(\nabla\phi) \frac{\partial\Gamma(\nabla\phi)}{\partial\nabla\phi} \otimes \nabla \phi, \nonumber \label{str2}
\end{equation}
where $p$ is the pressure, and $\nu$ is the viscosity which can be dependent on $\phi$ and $T$. $\nabla_{\text{s}}\mathbf{u}:=\frac{1}{2}(\nabla\mathbf{u}+\nabla\mathbf{u}^\top)$ is the deformation gradient, and $\mathbf{I}$ is identity tensor. Moreover, $\pmb{\sigma}_\mathrm{K}$ is the Korteweg stress tensor, which reflects the effect of interface tension on the fluid, i.e, capillary effect \cite{yang2020non, anderson1998diffuse}. For a simple isotropic interface energy, $\pmb{\sigma}_\mathrm{K}$ reduces to the well-known form $\pmb{\sigma}_\mathrm{K} = T\kappa_\phi\nabla \phi \otimes \nabla \phi$. Also, we note that for a more general anisotropic interface energy, $\pmb{\sigma}_\mathrm{K}$ becomes non-symmetric, thus breaking the angular momentum conservation, cf. \cite{Levitas2015}. Hence, to be precise, every matrix divergence is taken column-wise.

\noindent Following Eqs.~(\ref{entropy}) and (\ref{fed}), we can define the expressions ${\delta S}/{\delta \phi} = -(1/T){\delta F}/{\delta \phi}$, ${\delta S}/{\delta e} = 1/T$. Moreover, taking into account Eqs.~(\ref{fed}), (\ref{ns}) - (\ref{str1}), the general governing equations of the model can be defined as
\begin{align}
    \frac{\mathrm{D} \phi}{\mathrm{D} t} &= -\mathbf{L}_{\phi\phi}\frac{\mu}{T} + \mathbf{L}_{\phi e}\cdot\nabla\frac{1}{T}\label{ac}, \\
    \frac{\mathrm{D} e}{\mathrm{D} t} &= \nabla\cdot\left(\mathbf{L}_{\phi e}\frac{\mu}{T} - \mathbf{L}_{ee} \nabla \frac{1}{T}\right) + \nu\nabla_{\text{s}} \mathbf{u} : \nabla\mathbf{u} - \pmb{\sigma}_\mathrm{K} : \nabla\mathbf{u}, \label{ener} \\
     \varrho\frac{\mathrm{D} \mathbf{u}}{\mathrm{D} t} &= -\nabla p + \nabla \cdot (\nu(\phi) \nabla_{\text{s}} \mathbf{u}) - \nabla \cdot \pmb{\sigma}_\mathrm{K} + \mathbf{b}, \nonumber \\ 
     \nabla \cdot \mathbf{u} &= 0, \label{ns2}
\end{align}
with
\begin{equation}
    \frac{\mu}{T} = -\nabla\cdot\left(\kappa_\phi\Gamma(\nabla\phi)\frac{\partial \Gamma(\nabla\phi)}{\partial\nabla\phi}\right) + \frac{1}{T}\frac{\partial f}{\partial \phi}. \nonumber \label{mu}
\end{equation}
Here, we introduced for simplicity the term ${\mu}/{T} = (1/T){\delta F}/{\delta \phi}$ and further note that the pressure contribution to the internal energy is dropped out due to divergence-freedom. It is worth noting that an additional driving force, $\nabla \cdot \pmb{\sigma}_\mathrm{K}$, influences melt flow dynamics in \eqref{ns2}. This driving force is absent in most conventional phase-field models of solidification coupled with melt flow dynamics \cite{andersson2002phase, wang2021phase, sakane2021phase}. It arises from the Korteweg stress, which originates from the coupling between the phase-field and melt flow, and is associated with capillary effects \cite{anderson1998diffuse}. For a simple isotropic case, $\pmb{\sigma}_\mathrm{K}$ depends on $\kappa_\phi$, a phase-field parameter related to interface energy and width, meaning that $\nabla \cdot \pmb{\sigma}_\mathrm{K}$ represents a melt flow driving force due to variations in interface energy.
Moreover, since ${\sigma}_\mathrm{K}$ is temperature-dependent, $\nabla \cdot \pmb{\sigma}_\mathrm{K}$ also accounts for thermal capillary effects caused by temperature gradients at the liquid-solid interface. The explicit derivation and formulation of $\nabla \cdot \pmb{\sigma}_\mathrm{K}$ employed in this work are provided in the Appendix. Also, we consider the viscosity $\nu$ to be dependent on phase-field variable $\phi$. It is defined to interpolate between the viscosity of the solid, $\nu_\mathrm{s}$, and that of the liquid, $\nu_\mathrm{l}$, across the diffusive interface. This is typically referred to as the variable viscosity model, where a considerably high value of $\nu_\mathrm{s}/\nu_\mathrm{l}$ is assumed \cite{tonhardt1998phase, nestler2000phase}. The effects of different viscosity interpolations on the no-slip boundary condition at the interface are further investigated in Section~\ref{visco_inter}.

\noindent Furthermore, we note that the model equations system conserves the total energy and momentum (assuming no external body forces), i.e
\begin{align}
  \frac{\mathrm{d}}{\mathrm{d}t} &\int_\Omega e_{tot}(T,\phi,\mathbf{u})\mathrm{d}\Omega = 0, \qquad \qquad \frac{\mathrm{d}}{\mathrm{d}t} \int_\Omega \varrho \mathbf{u}\mathrm{d}\Omega = 0.
\end{align}
In addition, the entropy production, defined as 
\begin{align}
  \frac{\mathrm{d}}{\mathrm{d}t} &\int_\Omega s(T,\phi)\mathrm{d}\Omega  = \int_\Omega \begin{pmatrix}
      \frac{\mu}{T} \\ \nabla\frac{1}{T}
  \end{pmatrix}\cdot \begin{pmatrix}
      \mathbf{L}_{\phi\phi} & -\mathbf{L}_{\phi e}^\top \\ -\mathbf{L}_{\phi e} & \mathbf{L}_{ee}
  \end{pmatrix}\begin{pmatrix}
      \frac{\mu}{T} \\ \nabla\frac{1}{T}
  \end{pmatrix} + \frac{\nu(\phi)}{T}\nabla_{\text{s}} \mathbf{u}:\nabla_{\text{s}} \mathbf{u} \mathrm{d}\Omega\geq 0,
\end{align}
is guaranteed to be non-negative if the mobility matrix is taken semi-definite positive and the viscosity is non-negative. Specifically, this requires $\mathbf{L}_{\phi\phi}$ to be non-negative, $\mathbf{L}_{ee}$ to be positive semi-definite and $\det(\mathbf{L}_{\phi\phi}\mathbf{L}_{ee}- \mathbf{L}_{\phi e}\otimes\mathbf{L}_{\phi e})$ to be non-negative. Moreover, in the subsequent discussions and analyses, the cross-coupling terms between $\phi$ and $e$ in Eqs.~(\ref{ac}) and (\ref{ener}) are neglected, i.e. $\mathbf{L}_{\phi e}=0$. These terms are typically associated with antitrapping terms defined to eliminate artificial interface effects at the diffusive interface \cite{brener2012kinetic, boussinot2013interface, oyedeji2023variational}. Consequently, the phase-field mobility is taken as $\mathbf{L}_{\phi \phi}\equiv L_{\phi}$, and the energy mobility defined as $\mathbf{L}_{ee}=kT^2\mathbf{I}$, where $k$ is the thermal conductivity consistent Fourier's law.

\noindent In this work, we investigate solidification processes considering both isotropic and anisotropic interface energy. For systems with simple isotropic interface energy, the term $\kappa_\phi\Gamma^2(\nabla \phi)$ is simply taken as $\kappa_\phi\Gamma^2=\kappa_\phi|\nabla\phi|^2$. For anisotropic interface energy in dendrite simulations, we follow the approach proposed in Ref. \cite{karma1998quantitative}, taking $\Gamma^2(\nabla\phi) = \delta^2(\mathbf{n})|\nabla\phi|^2$ and ${L}_\phi(\mathbf{n})$ to be dependent on the interface normal vector $\mathbf{n}:={\nabla\phi}/{|\nabla\phi|}$:
\begin{equation}
    \delta(\mathbf{n}) = \delta_0 a(\mathbf{n}),\qquad {L}_\phi(\mathbf{n}) = \left(\tau_0 a^2(\mathbf{n})\right)^{-1}, \label{an1}
\end{equation}
where $\delta_0>0$ and $\tau_0>0$ are amplitude constants. The anisotropy factor $a(\mathbf{n})$ can be defined for a $m$-fold symmetry as $a(\mathbf{n}):=(1+\epsilon\cos(m\theta))$, where $\theta=\mathrm{atan}(\frac{\partial_y \mathbf{n}}{\partial_x \mathbf{n}})$ \cite{kobayashi1993modeling}. For the fourfold anisotropy ($m=4$) employed in this work, the explicit expression is given as
\begin{equation}
    a(\mathbf{n}) = (1 -3\epsilon)\left[1 + \frac{4\epsilon}{1-3\epsilon}\frac{\sum_i (\partial \phi/\partial x_i)^4}{|\nabla \phi|^4} \right],  \label{an2}
\end{equation}
where $\epsilon$ is the strength of anisotropy.




\section{Results and Discussion} \label{model_result}
\subsection{Model Verification} \label{model_verify}
\begin{figure}[!h]
\centering
\includegraphics[width=\textwidth]{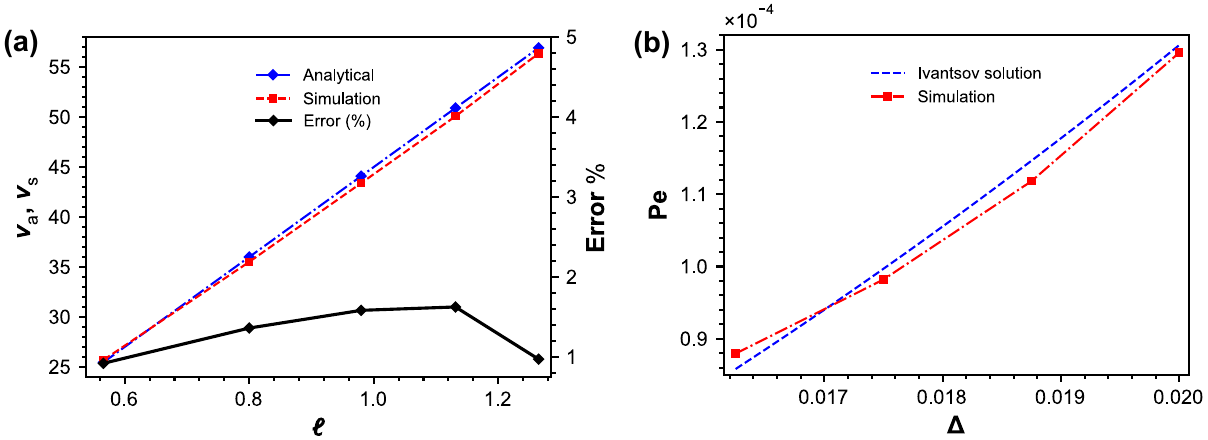}
\caption {(a) Comparison of analytical predictions and simulated interface velocities for various interface widths during 1D planar solidification. Here, $v_a$ represents the analytically predicted velocity, while $v_s$ denotes the simulated velocity; (b) Comparison of the 2D Ivantsov solution for the Peclet number ($Pe$) with the Peclet number obtained from simulation results, both plotted as functions of undercooling ($\Delta$).} 
\label{fig:plane_mode_verify}
\end{figure}

In this section, we numerically verify the accuracy of the presented phase-field model by simulating the solidification of a planar solid–liquid interface and dendritic growth, both without melt flow dynamics. For the planar interface, we consider an analysis with a one-dimensional (1D) profile of the phase-field variable $\phi$, defined as $\phi_0 = 0.5[1 - \tanh(\sqrt{H}/\sqrt{2\kappa})]$. The simulated interface velocity, $v_\mathrm{s}$, is compared with the analytical steady-state velocity, $v_\mathrm{a} = M f_\mathrm{ht}$, where $M = 6L_\phi \sqrt{\kappa/2H}$ \cite{takaki2014phase}. Fig.~\ref{fig:plane_mode_verify}a shows the comparison between the analytical predictions and the simulated interface velocity for a constant driving force, $f_\mathrm{ht}$, across different interface widths. The results indicate minimal differences between the predicted simulated velocities and the sharp-interface analytical solutions, with an error difference of $< 2\%$ for all interface widths considered. 

Furthermore, we conduct simulations of dendritic growth and compare the results with the classical Ivantsov solution \cite{ivantsov1947temperature}. The Ivantsov solution describes the relationship between rescaled undercooling, defined as $\Delta = c_p(T - T_M)/\mathcal{L}$, and the growth Peclet number, $Pe = VR/2D$, for diffusion-controlled solidification, where $V$ represents the dendrite tip velocity, $R$ is the radius of the dendrite tip, and $D = k/c_p$ is the thermal diffusivity. For two-dimensional (2D) dendritic growth, the Ivantsov function is given by
\begin{equation}
    \Delta = \sqrt{\pi Pe} \exp(Pe) \mathrm{erfc}(\sqrt{Pe}).
    \label{pe_eq}
\end{equation}
The schematic of the 2D simulation domain for the dendritic growth simulation is shown in Fig.~\ref{fig:sche_solid}a. The domain size is $L_\mathrm{x} = L_\mathrm{y} $ with an initial solid seed of radius $r$.  The simulation parameters employed are given in Table~\ref{table1}. Fig.~\ref{fig:plane_mode_verify}b compares the simulated growth Peclet numbers with the Ivantsov prediction from \eqref{pe_eq}. The results show that $Pe$ increases with increasing undercooling $\Delta$, and the simulated $Pe$ exhibit good agreement with the Ivantsov solution, further validating the model’s accuracy in capturing dendritic growth dynamics.


\begin{table*}
\centering
\caption{\label{table1} Set of dimensionless quantities and simulation parameters employed for dendritic growth simulations} \setlength{\tabcolsep}{6mm}{
\begin{tabular}{lllllllllll}
\toprule
         $L_x$   & $L_y$ & $r$ & $\delta_0$ & $\tau_0$ & $\epsilon$ & $c_p$ & $\mathcal{L}$ & $\nu_\mathrm{s}$ & $\nu_\mathrm{l}$ & k\\
            \hline
25  &  25  &   0.2 &  0.01   &  0.01 & 0.05 & 1 & 20 & 1000 & 1 & 1\\
\bottomrule
\end{tabular}}
\end{table*}

\subsection{Thermal Capillary Effect} \label{thermal_cap}
\begin{figure}[!h]
\centering
\includegraphics[width=\textwidth]{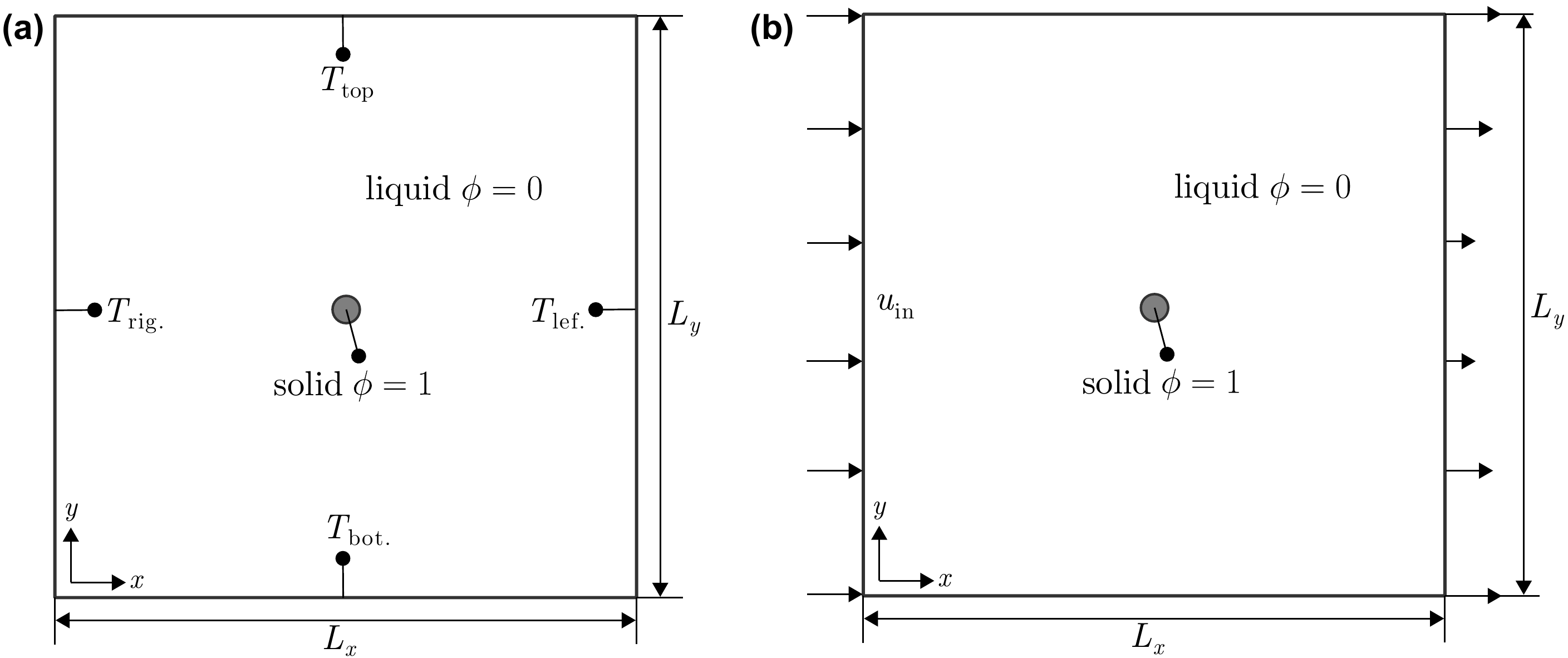}
\caption{Schematic illustrations of the simulation setups for dendritic growth, showing the specified initial and boundary conditions for (a) conventional dendrite growth without imposed flow; (b) simulations with imposed forced flow. Note that the temperature boundary conditions from (a) also apply to (b). \label{fig:sche_solid}} 
\end{figure}

\begin{figure}[!h]
\centering
\includegraphics[width=\textwidth]{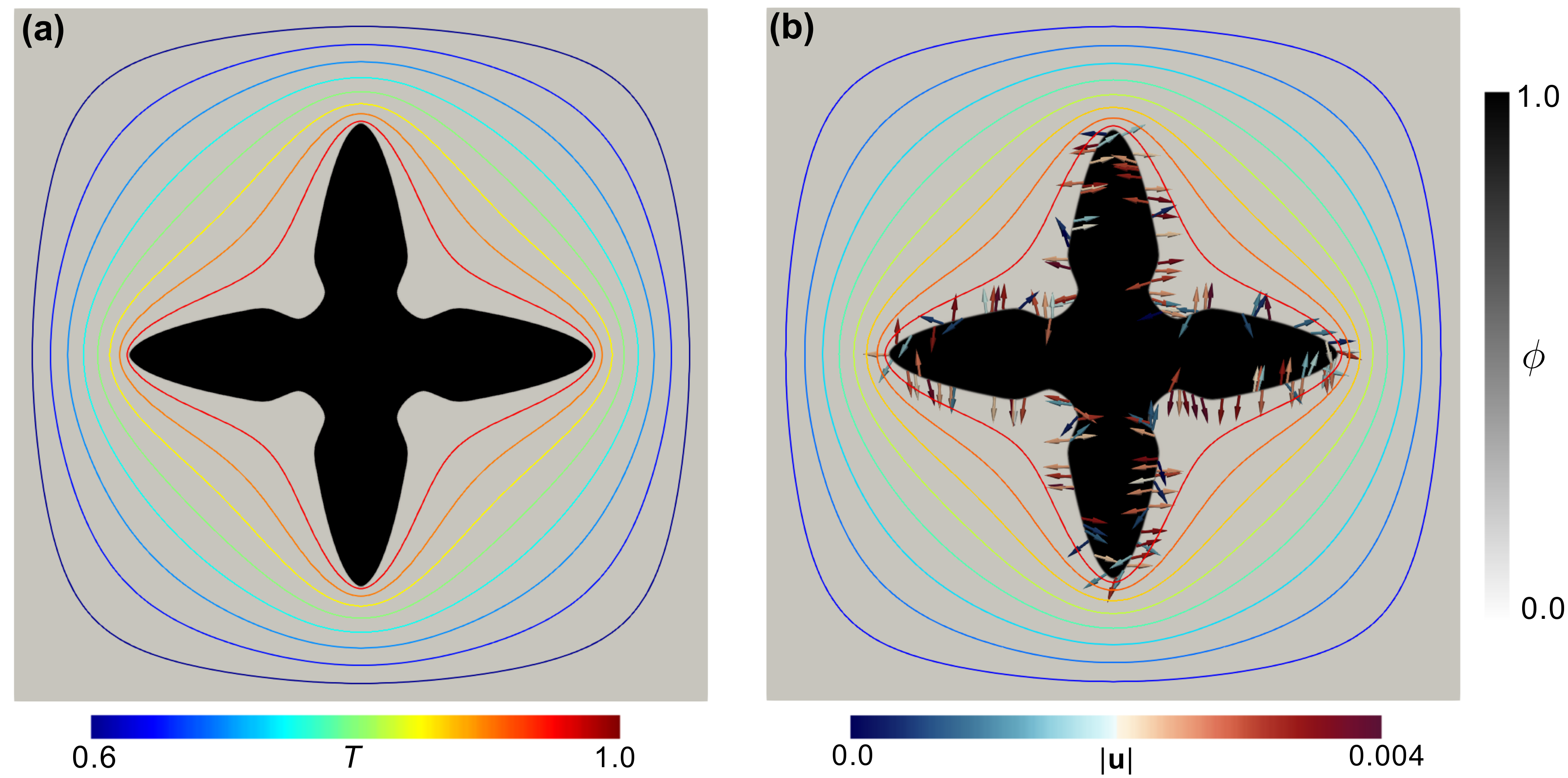}
\caption{Simulated dendritic morphologies, temperature isolines, and melt flow velocity fields for (a) the case without thermal capillary effect; (b) the case with thermal capillary effect. \label{fig:therm_cap}} 
\end{figure}

\begin{figure}[!h]
\centering
\includegraphics[width=\textwidth]{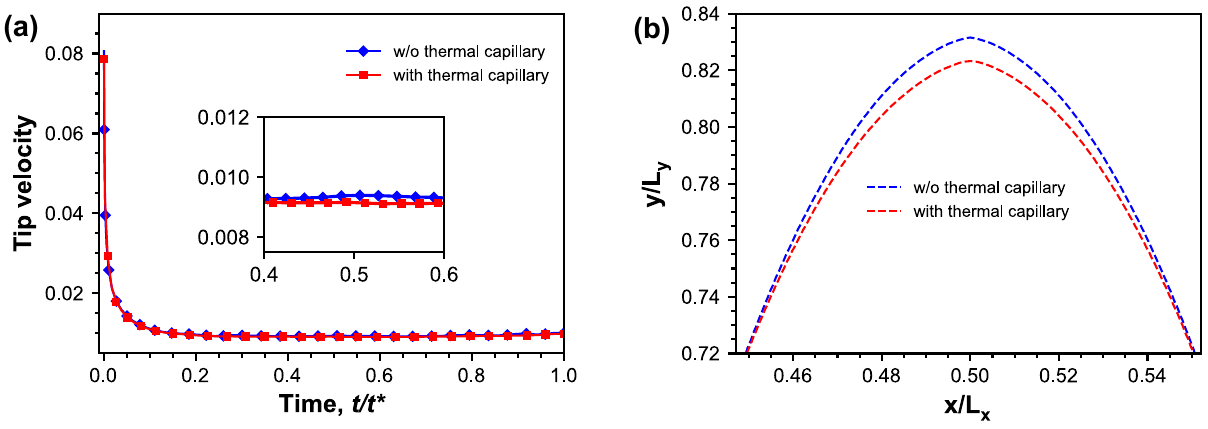}
\caption{(a) Temporal evolution of the dendrite tip velocity comparing simulations with and without thermal capillary effects. $t^* = 500$ units; (b) Comparison of the final north dendrite tip positions for both cases. \label{fig:therm_cap_plots}} 
\end{figure}

To investigate the effect of the driving force $\nabla\cdot\pmb{\sigma}_\mathrm{K}$ in the melt flow dynamics equation \eqref{ns2}, which represents thermal capillary effect, we performed simulations of dendritic growth coupled with melt flow. The simulation setup follows the schematic illustrated in Fig.~\ref{fig:sche_solid}a, employing simulation parameters outlined in Table~\ref{table1}. The liquid temperature is initially set uniformly to $T = 0.60$, i.e $T_\mathrm{top} = T_\mathrm{bot.} = T_\mathrm{rig.} = T_\mathrm{left.} = 0.60$ with no body force ($\mathbf{b}=0$). The solid temperature is initially set to the melting temperature, $T_m  = 1$, and the melt flow velocity is set to no-slip at all boundaries, i.e. $\mathbf{u}\vert_{\partial\Omega}=\mathbf{0}$. 
Fig.~\ref{fig:therm_cap} compares dendritic morphologies, temperature isolines, and melt flow velocity profiles for simulations with and without the thermal capillary effect, that is, with and without inclusion of the term $\nabla\cdot\pmb{\sigma}_\mathrm{K}$ in \eqref{ns2}. For the case without the thermal capillary effect (Fig.~\ref{fig:therm_cap}a), dendritic growth occurs regularly, and no melt flow is generated. In contrast, when the thermal capillary effect is included (Fig.~\ref{fig:therm_cap}b), melt flow is induced around the solid–liquid interface during dendritic growth. To further examine the influence of the induced melt flow on dendrite evolution, we analyze the temporal evolution of the dendrite tip velocity. Fig.~\ref{fig:therm_cap_plots}a shows that the presence of melt flow due to the thermal capillary effect results in a slightly reduced dendrite tip velocity compared to the case without it. This reduction in growth velocity leads to a shorter dendrite tip, as evident by the comparison of the north dendrite tips in Fig.~\ref{fig:therm_cap_plots}b.

\begin{figure}[!h]
\centering
\includegraphics[width=\textwidth]{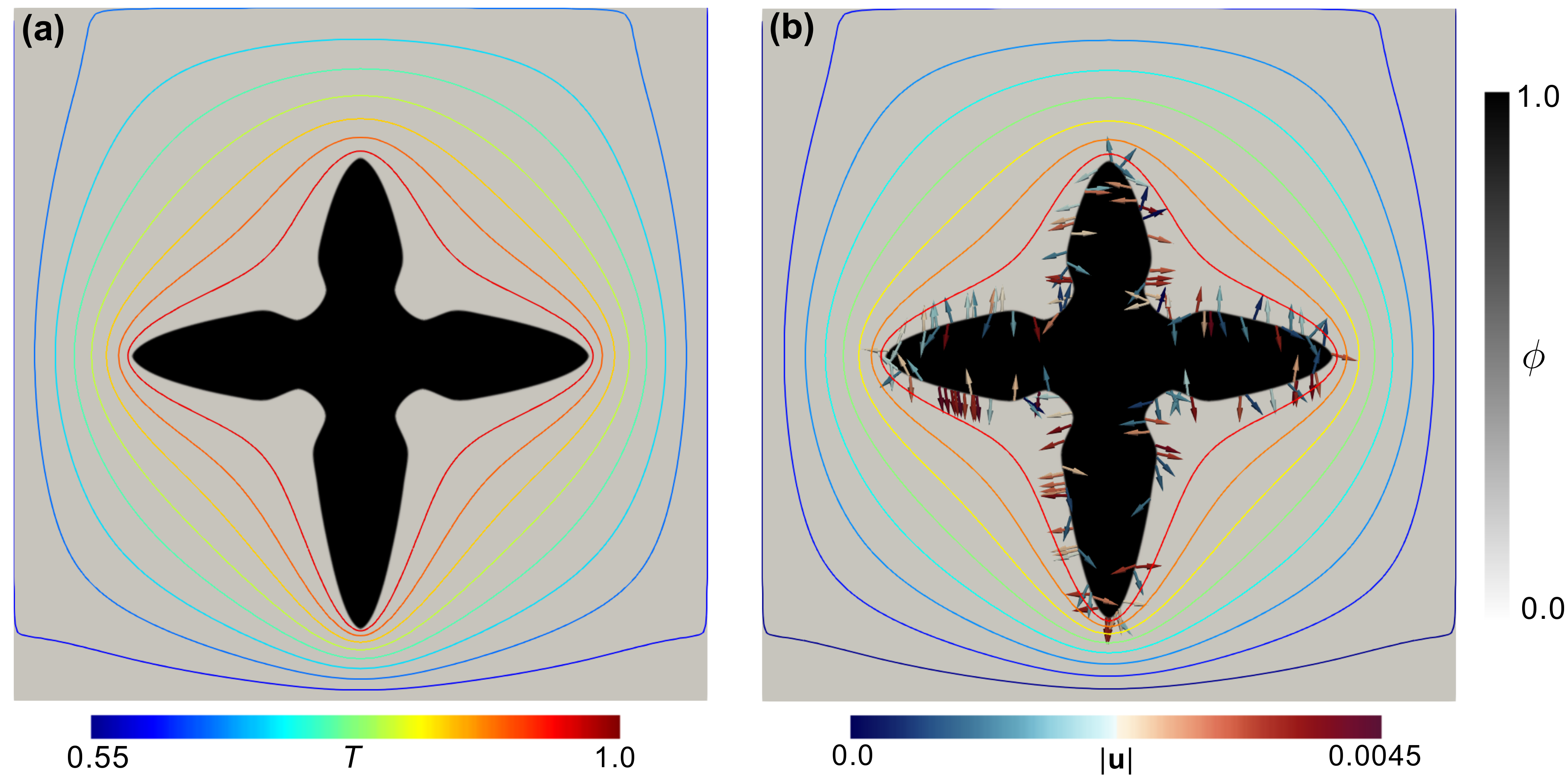}
\caption{Simulated dendritic morphologies, temperature isolines, and melt flow velocity fields for a case with an imposed temperature gradient: (a) without thermal capillary effect; (b) with thermal capillary effects. \label{fig:therm_cap_grad}} 
\end{figure}

\begin{figure}[!h]
\centering
\includegraphics[width=\textwidth]{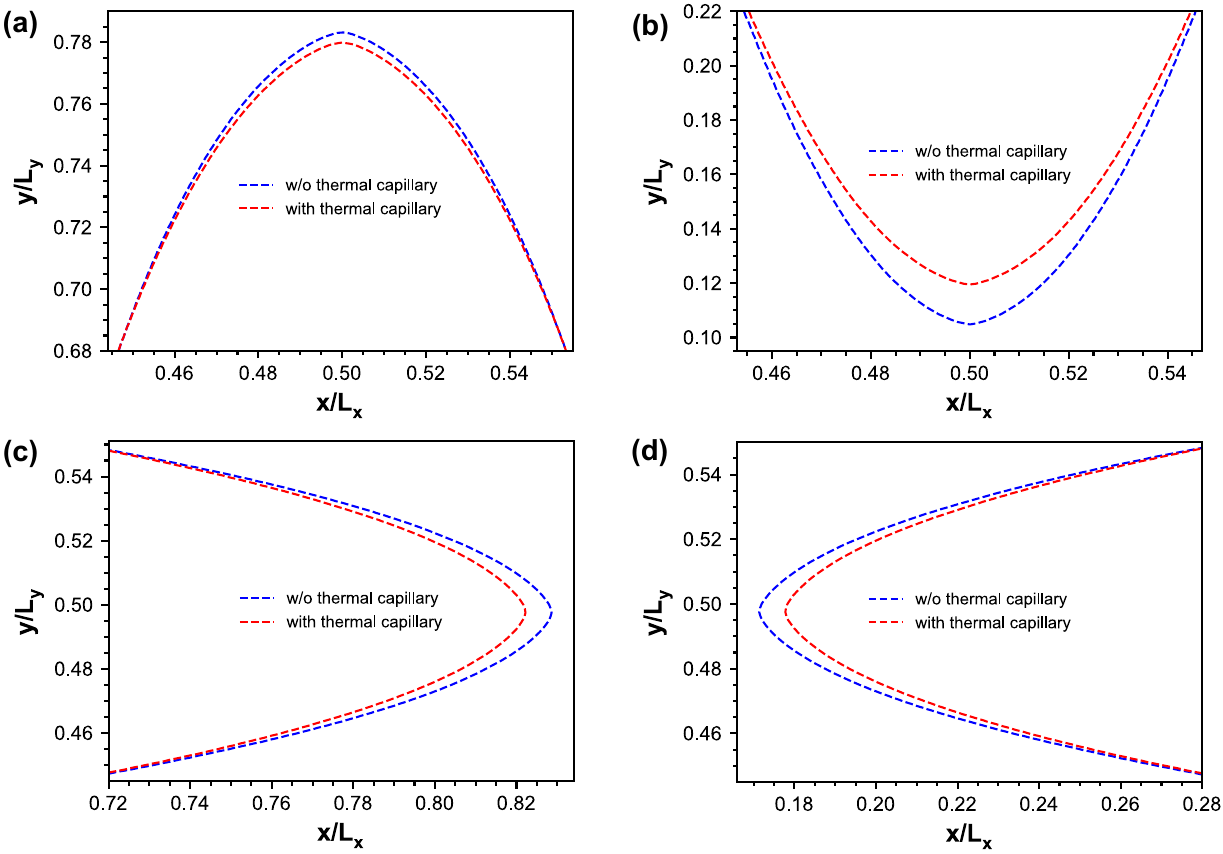}
\caption{Comparison of simulated dendrite tip positions under a temperature gradient for cases with and without thermal capillary effects: (a) north tip, (b) south tip, (c) east tip, and (d) west tip. \label{fig:therm_cap_grad_plts}} 
\end{figure}

Additionally, we consider a second simulation scenario based on the same schematic (Fig.~\ref{fig:sche_solid}a) but with a temperature gradient imposed in the liquid. In this case, the temperature at the right and left boundaries of the simulation domain are kept at $T_\mathrm{rig.} = T_\mathrm{left.} = 0.60$, while the temperature at the top and bottom boundaries are set at $T_\mathrm{top} = 0.65$ and $T_\mathrm{bot.} = 0.55$, respectively. Simulations are again performed both with and without the thermal capillary effect. Fig.~\ref{fig:therm_cap_grad} presents the resulting dendrite morphologies, temperature fields, and melt flow velocity profiles for both cases. In both scenarios, the east and west dendrite tips exhibit similar growth velocities due to symmetric undercooling conditions. However, the north and south tips grow at different rates as a result of the vertical temperature gradient. When the thermal capillary effect is included (Fig.~\ref{fig:therm_cap_grad}b), melt flow is again induced near the solid–liquid interface. As observed previously, this induced flow impacts dendrite growth dynamics. Fig.~\ref{fig:therm_cap_grad_plts} compares the final positions of the north, south, east, and west dendrite tips for both cases. The results indicate that the induced melt flow consistently reduces dendrite tip velocities, resulting in shorter dendrite arms in all directions when the thermal capillary effect is considered.

\subsection{Effect of Forced Flow}
\begin{figure}[!h]
\centering
\includegraphics[width=\textwidth]{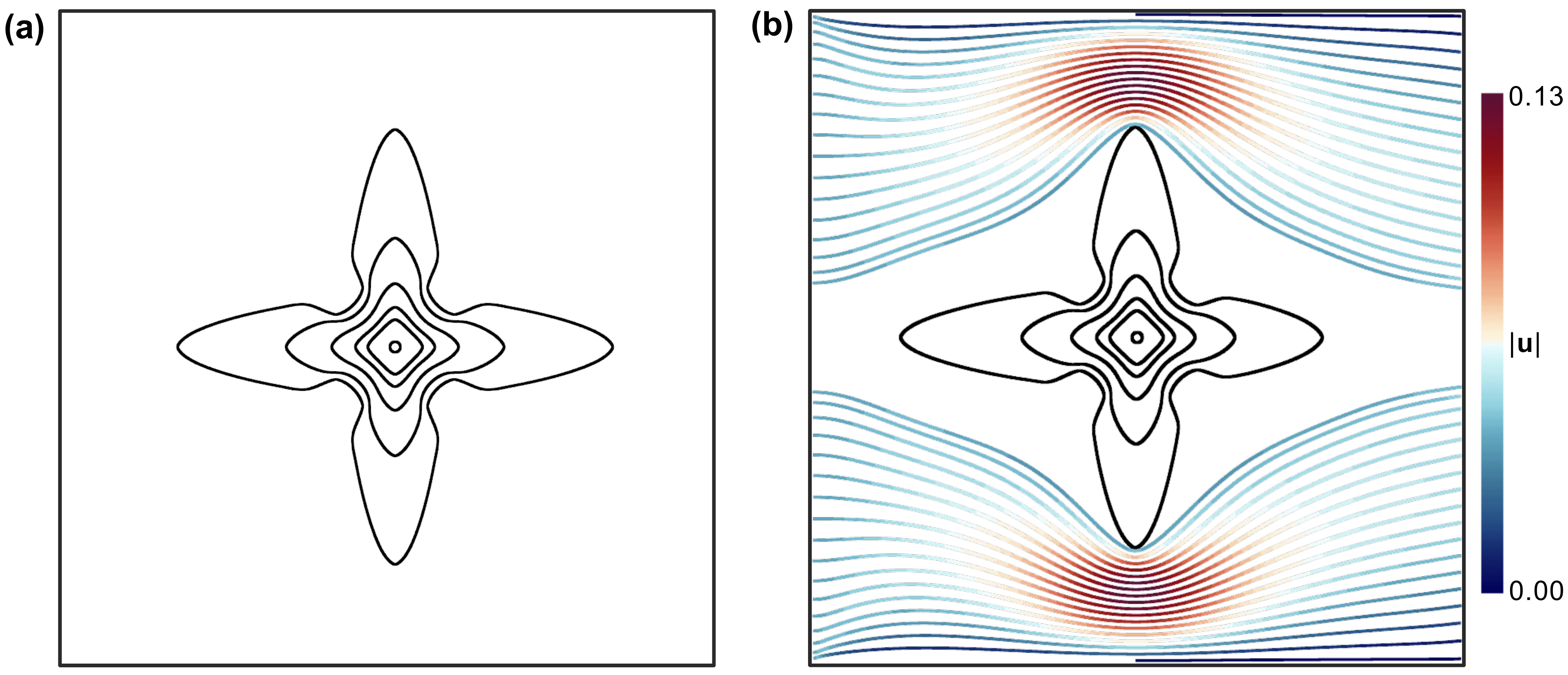}
\caption{Simulated solid-liquid interfaces at $t/t^* = 0, 0.0625, 0.125, 0.25, 0.5, 1$ for (a) dendritic growth without inlet flow ($u_\mathrm{in} = 0$); (b) dendritic growth with imposed inlet flow ($u_\mathrm{in} = 0.04$). Flow velocity streamlines are also indicated. \label{fig:dendr_struc_flow}} 
\end{figure}

\begin{figure}[!h]
\centering
\includegraphics[width=\textwidth]{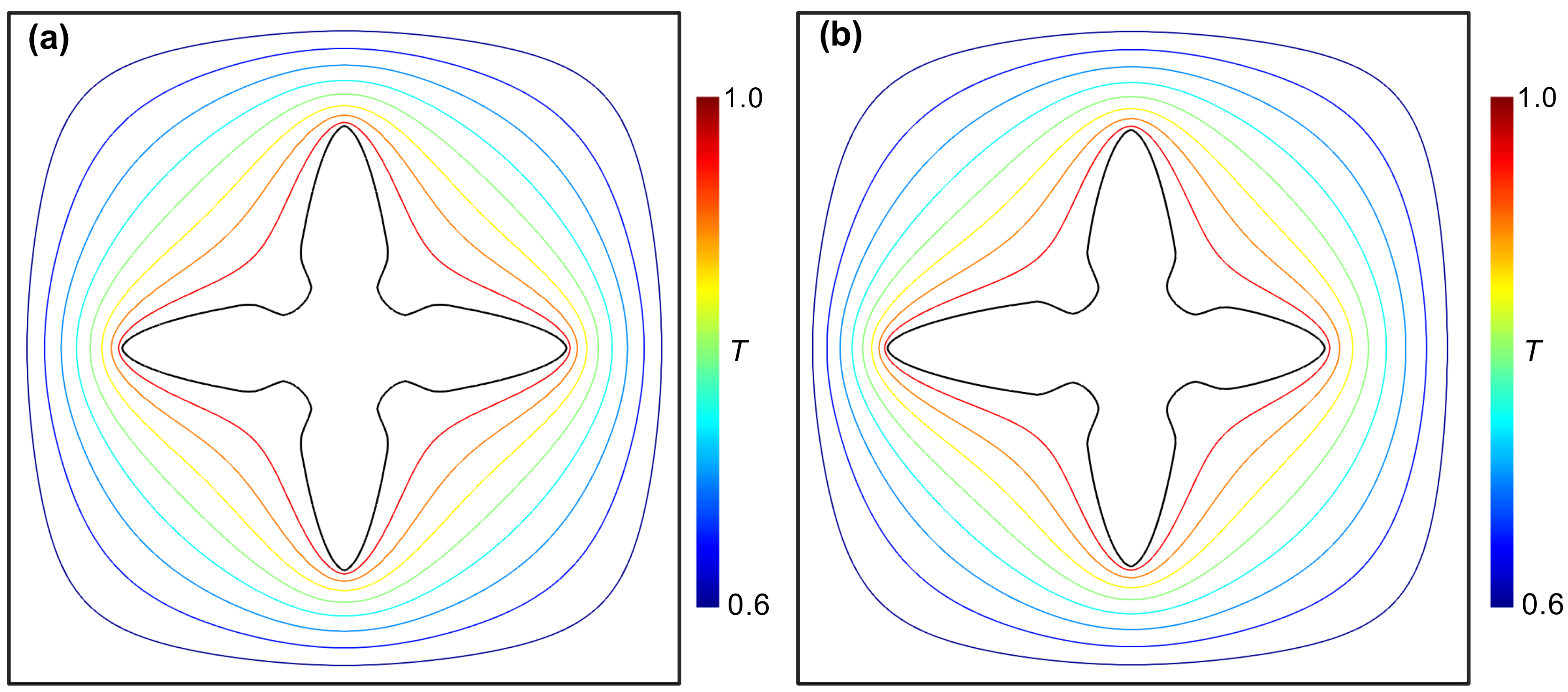}
\caption{Simulated solid-liquid interface and temperature isolines for (a) dendritic growth without inlet flow ($u_\mathrm{in} = 0$); (b) dendritic growth with imposed inlet flow ($u_\mathrm{in} = 0.4$).\label{fig:dendr_flow_temp}} 
\end{figure}

\begin{figure}[!h]
\centering
\includegraphics[width=\textwidth]{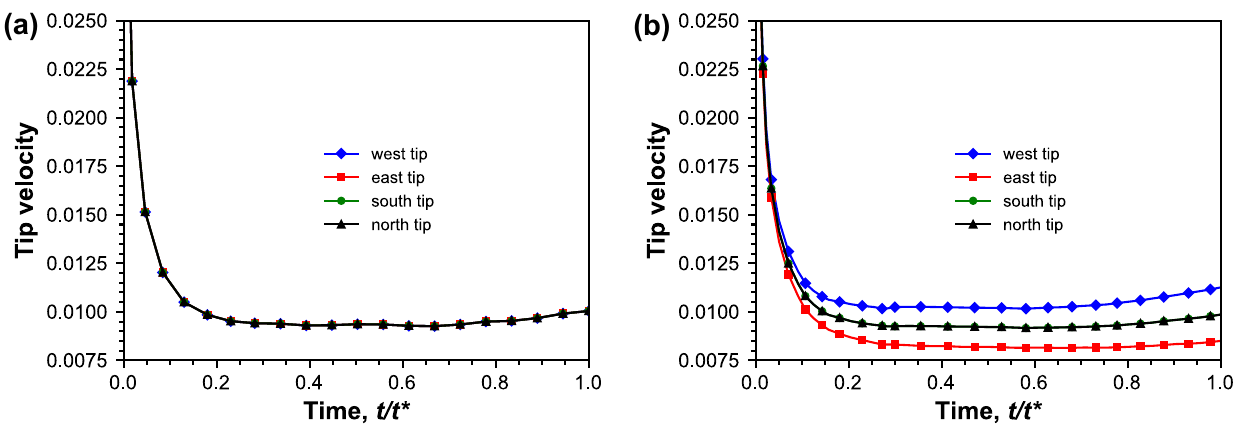}
\caption{Temporal evolution of the dendrite tips velocity for (a) dendritic growth without inlet flow ($u_\mathrm{in} = 0$); (b) dendritic growth with imposed inlet flow ($u_\mathrm{in} = 0.4$). $t^* = 500$ units. \label{fig:dendr_flow_evol}} 
\end{figure}

\begin{figure}[!h]
\centering
\includegraphics[width=\textwidth]{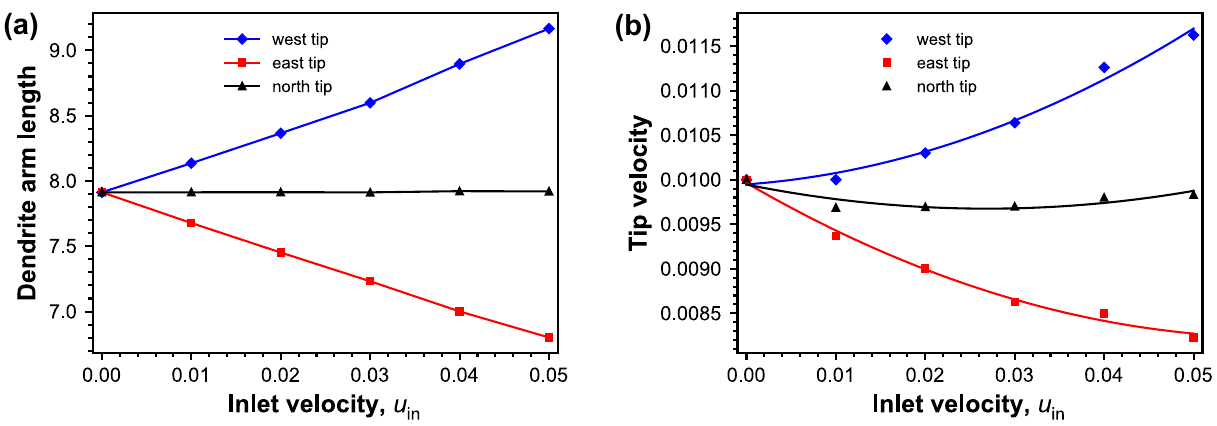}
\caption{Simulated (a) dendrite arm lengths and (b) tip growth velocities for west, east, and north tips for different values of inlet velocities $u_\mathrm{in}$. Solid lines represent fitted trends.\label{fig:dendr_diff_vel}} 
\end{figure}

In this section, we investigate the effect of forced flow on the solidification process by simulating dendritic growth subjected to an imposed inlet velocity $u_\mathrm{in}$ along the $x$-axis. The simulation schematic is given in Fig.~\ref{fig:sche_solid}b, employing simulation parameters outlined in Table~\ref{table1}. The inlet flow velocity is set to $u_\mathrm{in} = 0.4$, and the liquid temperature is initialized uniformly at $T = 0.60$ i.e $T_\mathrm{top} = T_\mathrm{bot.} = T_\mathrm{rig.} = T_\mathrm{left.} = 0.60$. The solid temperature is initially set to the melting temperature, $T_m  = 1$, and the melt flow velocity is set to no-slip at the top and bottom boundaries. At the outflow boundary we employ do-nothing boundary conditions.
Fig.~\ref{fig:dendr_struc_flow} presents the temporal evolution of the solid-liquid interface ($\phi=0.5$) for two cases: one without inlet flow ($u_\mathrm{in} = 0$), and one with inlet flow ($u_\mathrm{in} = 0.4$). The corresponding flow velocity field for the case with inlet flow is shown in Fig.~\ref{fig:dendr_struc_flow}b. The associated temperature isolines for both cases are presented in Fig.~\ref{fig:dendr_flow_temp}. The results demonstrate that the introduction of forced flow at the inlet breaks the symmetry of dendritic growth across the domain. Specifically, the west dendrite tip (i.e., on the upstream side of the flow) exhibits accelerated growth, while the east tip (on the downstream side) is suppressed. Meanwhile, the north and south tips grow symmetrically and remain largely unaffected by the imposed inlet flow. This asymmetry in dendrite growth can be attributed to the non-uniform temperature distribution induced by the inlet flow, as shown in Fig.~\ref{fig:dendr_flow_temp}. For the no-flow case ($u_\mathrm{in} = 0$), the temperature isolines are uniformly distributed around all dendrite tips. However, for the forced flow case ($u_\mathrm{in} = 0.4$), there is a variation of temperature distribution around the dendrite tips due to convective heat transfer. The upstream region experiences a larger temperature gradient, promoting faster dendritic growth at the west tip. In contrast, the downstream region exhibits a smaller temperature gradient, leading to reduced growth at the east tip. Fig.~\ref{fig:dendr_flow_evol} compares the time evolution of the dendrite tip velocities for both cases. In the absence of inlet flow, all tips grow at similar velocities, indicating uniform solidification. However, with forced flow, the west tip velocity increases, while the east tip velocity decreases, due to the asymmetric influence of flow. The north and south tip velocities remain largely unchanged, due to the symmetric temperature and flow fields in those directions. Fig.~\ref{fig:dendr_diff_vel} shows the dendrite arm lengths and tip velocities at the simulation end time $t^*$ for different values of $u_\mathrm{in}$. The results indicate that increasing $u_\mathrm{in}$ leads to an increase in the west tip length and velocity, and a corresponding decrease for the east tip. The north tip experiences only minor variations in length and velocity, showing the directional influence of the forced flow on dendritic growth.




\subsection{Viscosity Interpolation Comparison} \label{visco_inter}
\begin{figure}[!h]
\centering
\includegraphics[width=\textwidth]{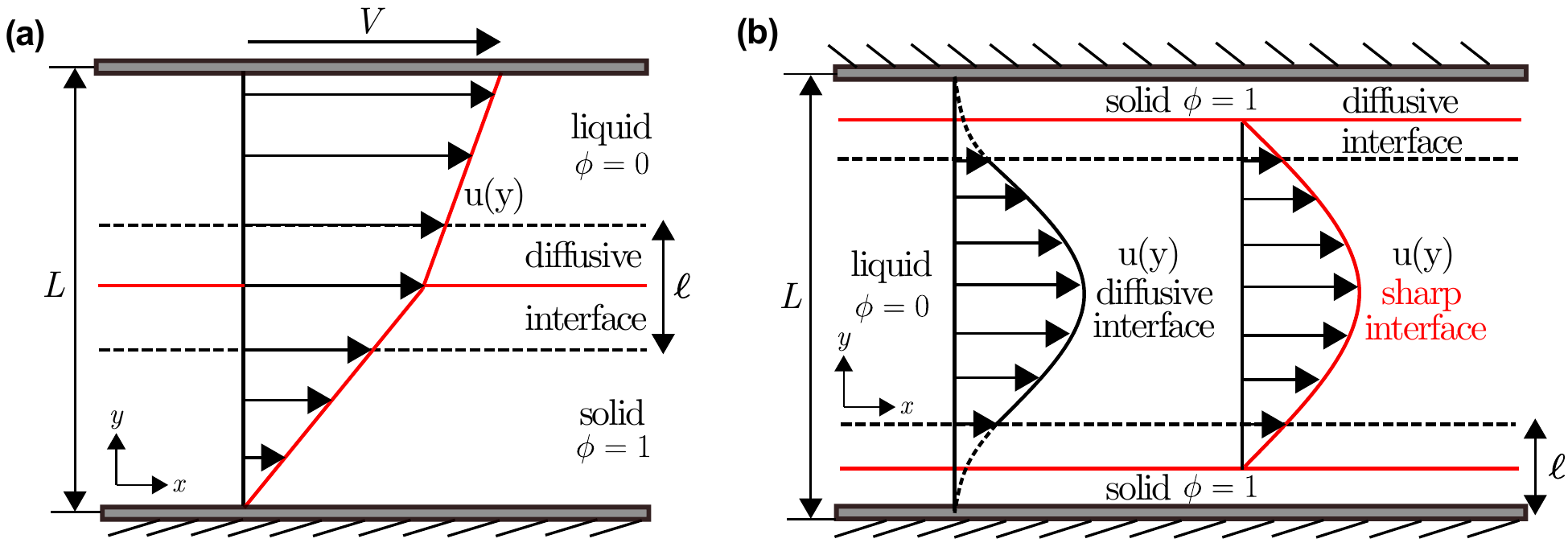}
\caption {Schematic illustrations of the simulation setups: (a) Couette flow with a stationary lower wall and an upper wall moving at velocity $V = 10$, separated by a distance $L = 50$; (b) Poiseuille flow between two stationary walls separated by a distance $L = 10$.} 
\label{fig:planar}
\end{figure}

Here, we analyze the impact of the viscosity interpolation scheme on melt flow simulations by comparing the inverse and direct interpolation schemes in Eqs.~(\ref{ii}) and (\ref{di}), respectively. 
\begin{align}
    \frac{1}{\nu(\phi)} = \frac{h(\phi)}{\nu_\mathrm{s}} + \frac{1-h(\phi)}{\nu_\mathrm{l}},
    \label{ii} \\
    \nu(\phi) = h(\phi) \nu_\mathrm{s} + (1-h(\phi))\nu_\mathrm{l}.
    \label{di}
\end{align}
Specifically, we examine the ability of each scheme to enforce the no-slip boundary condition ($\mathbf{u} = 0$ in solid) in numerical simulations of solidification coupled with melt flow. Although the viscosities of the solid ($\nu_\mathrm{s}$) and liquid ($\nu_\mathrm{l}$) are interpolated using the function, $h(\phi)$ employed in this work, the analysis is applicable to other arbitrary interpolation functions. To evaluate the accuracy of the schemes, we consider classical Couette and Poiseuille flow cases, with schematic representations shown in Fig.~\ref{fig:planar}. For simplicity, the cases are modeled under isothermal conditions with no applied pressure gradient. The numerically obtained velocity profiles $\mathbf{u}$ are compared with analytical sharp-interface solutions. In the Couette flow case (Fig.~\ref{fig:planar}a), a stationary interface aligned with the $x$ axis separates the solid ($\phi = 1$) and liquid ($\phi = 0$) phases. The plane at $y = 0$ (solid region) is fixed while the upper plane at $y = L$ (liquid region) moves with a constant velocity $V$ in the $x$ direction. The body force $\mathbf{b}$ is set to zero. Taking all these into account, the velocity profile derived from \eqref{ns2} takes the form $u(y) = {Ay}/{\nu(\phi)} + B$ where $A$ and $B$ are integration constants determined from the boundary conditions. 

\begin{figure}[!h]
\centering
\includegraphics[width=\textwidth]{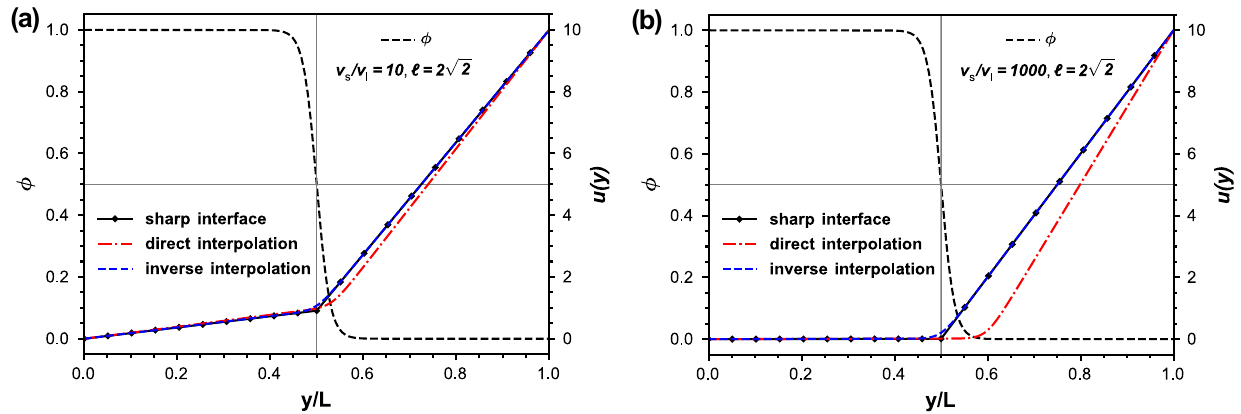}
\caption{Velocity profiles along the $y$-direction for Couette flow, comparing the analytical sharp-interface solution with numerically obtained profiles using both direct and inverse viscosity interpolation schemes for (a) $\nu_\mathrm{S}/\nu_\mathrm{L} = 10$ and (b) $\nu_\mathrm{S}/\nu_\mathrm{L} = 1000$.  \label{fig:diff_visco}} 
\end{figure}

\begin{figure}[!h]
\centering
\includegraphics[width=\textwidth]{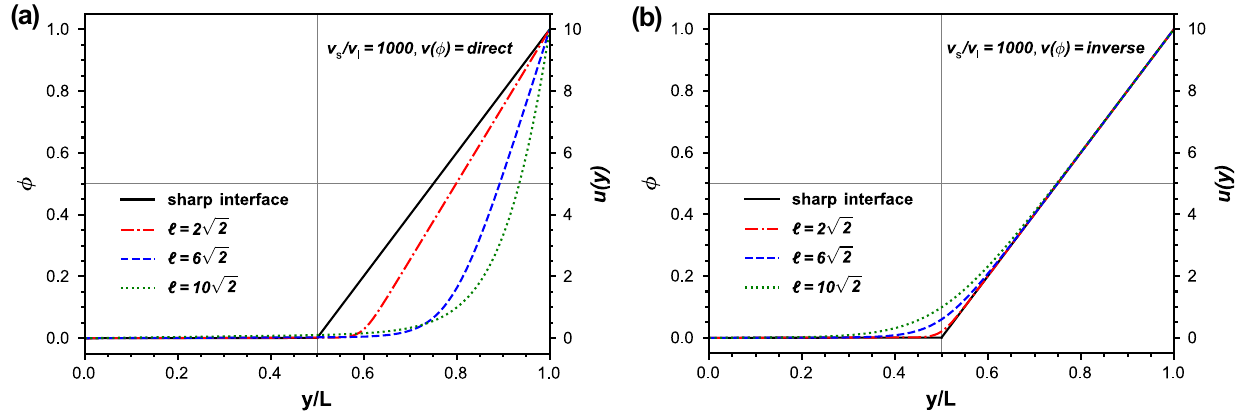}
\caption{Velocity profiles along the $y$-direction for Couette flow with a viscosity ratio of $\nu_\mathrm{S}/\nu_\mathrm{L} = 1000$ and varying interface width $\ell$. (a) Results using the direct viscosity interpolation from \eqref{di}; (b) results using the inverse interpolation from \eqref{ii}. Analytical sharp-interface profiles are shown for comparison. \label{fig:visco_width}} 
\end{figure}

\noindent Fig.~\ref{fig:diff_visco} presents the analytical and numerically obtained velocity profiles $u(y)$ for the viscosity ratios $\nu_\mathrm{S}/\nu_\mathrm{L} = 10$ and $1000$ using both inverse and direct viscosity interpolation schemes. The results show that the accuracy of the numerical solution is strongly dependent on the choice of interpolation function $\nu(\phi)$. When the direct interpolation (Eq.~(\ref{di})) is employed, significant deviations from the analytical profile are observed, particularly in the high-viscosity ratio regime, which is typical in solidification simulations. On the other hand, the inverse interpolation scheme (Eq.~(\ref{ii})) yields velocity profiles that closely match the sharp-interface solutions across both the solid and liquid domains, independent of the viscosity ratio. Further analysis, presented in Fig.~\ref{fig:visco_width} shows the influence of interface thickness $\ell$ on the velocity profile for $\nu_\mathrm{s}/\nu_\mathrm{l} = 1000$. Velocity profiles obtained numerically using both interpolation schemes are compared against the analytical solution. The results confirm that direct interpolation leads to an increasing deviation from the sharp-interface solution as $\ell$ increases. In contrast, inverse interpolation consistently recovers the analytical solution in both phases, independent of $\ell$, aligning with prior variable viscosity model analysis \cite{nestler2000phase}.

\begin{figure}[!h]
\centering
\includegraphics[width=\textwidth]{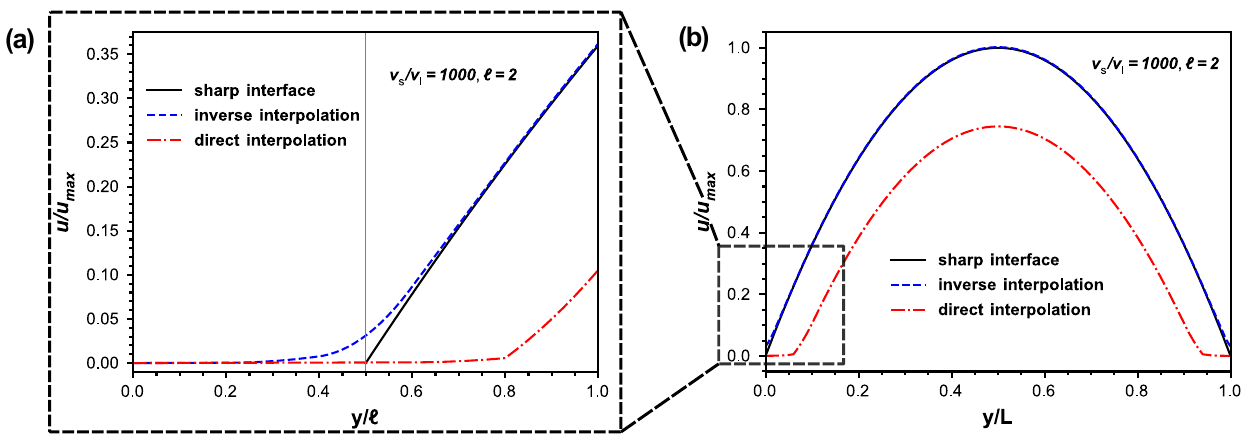}
\caption{(a) Velocity profiles along the $y$-direction for Poiseuille flow, comparing the analytical sharp-interface solution with numerical results obtained using both direct and inverse viscosity interpolation schemes; (b) Magnified view of the interface region in (a), highlighting differences around the interface. \label{fig:vel_pouise}} 
\end{figure}

\noindent Additionally, we examine the Poiseuille flow shown in Fig.~\ref{fig:planar}b. In this setup, two stationary solid walls enclose a liquid phase, with the solid-liquid interfaces remaining stationary. A body force $\mathbf{b}$ is applied in the direction of flow. The analytical velocity profile is obtained from \eqref{ns2} as  
\begin{equation}
    u(y) = 4u_\mathrm{max}\frac{y}{L}\bigg(1 - \frac{y}{L}\bigg) \label{1f},
\end{equation}
with $u_\mathrm{max} = {bL^2}/{8\nu(\phi)}$ \cite{subhedar2015modeling}. The analytical solutions from \eqref{1f} are shown in Fig.~\ref{fig:vel_pouise} for a viscosity ratio of $\nu_\mathrm{S}/\nu_\mathrm{L} = 1000$. Comparison with numerical results using both interpolation schemes shows consistent trends with those observed in Couette flow. The direct interpolation leads to significant deviations in the velocity profile within both phases, whereas the inverse interpolation reproduces the analytical solution accurately. The results presented above demonstrates the importance of the viscosity interpolation scheme in phase-field simulations involving melt flow. The inverse interpolation scheme offers excellent agreement with the sharp-interface solution, effectively enforcing the no-slip condition and ensuring accurate representation of melt flow dynamics.




\section{Conclusion} \label{model_conc}

In this work, we presented a thermodynamically consistent, non-isothermal phase-field model for solidification coupled with melt flow dynamics.  The model explicitly incorporates cross-coupling terms required for thermodynamic consistency, including the divergence of the Korteweg stress in the momentum equation, a term often neglected in existing phase-field-Navier-Stokes formulations. This stress not only ensures consistency with non-equilibrium thermodynamics but also accounts for thermal capillary effects. Numerical verification of the model against theoretical predictions for solidification-only scenarios, including planar interface evolution and dendritic growth, yielded excellent agreement. Furthermore, numerical simulations revealed that thermal capillary effects induce melt flow around the solid–liquid interface, in contrast with simulations where this effect was absent. This induced flow was shown to slightly reduce the dendrite tip growth velocity, resulting in a shorter dendrite. The effect of forced convection on dendritic growth was also investigated. The results demonstrated that forced convection redistributes the temperature field around the dendrite tips, leading to asymmetric dendrite growth in the flow direction, with greater asymmetry observed at higher flow velocities. Finally, we numerically examine the influence of different viscosity interpolation schemes on enforcing the no-slip boundary condition at the interface. The inverse interpolation scheme was found to reproduce the sharp-interface solution accurately, emphasizing its importance in phase-field melt flow coupled modeling and simulations.



\section*{Acknowledgement}

Authors acknowledge the financial support of German Science Foundation (DFG) in the Priority Program 2256 (SPP 2256, project number 441153493) and Collaborative Research Center Transregio 270 (CRC-TRR 270, project number 405553726, sub-projects A06, B13, Z-INF). The authors also greatly appreciate the access to the Lichtenberg II high-performance computer by the NHR Center NHR4CES@TUDa (funded by the German Federal Ministry of Education and Research and the Hessian Ministry of Science and Research, Art and Culture) and High-performance computer HoreKa by the NHR Center NHR@KIT
(funded by the German Federal Ministry of Education and Research and the Ministry of Science, Research and the Arts of Baden-Württemberg, partly funded by the DFG). The NHR4CES Resource Allocation Board allocates computing time on the HPC under the project “special00007”.


\section{Appendix: Reformulation of $\nabla\cdot\pmb{\sigma}_\mathrm{K}$} \label{append}

\noindent Here, we give an explicit reformulation of $\nabla\cdot\pmb{\sigma}_\mathrm{K}$ in the melt flow dynamics of \eqref{ns2}. Taking $\hat{\pmb{\sigma}} = \kappa_\phi\Gamma(\nabla\phi) \frac{\partial\Gamma(\nabla\phi)}{\partial\nabla\phi} \otimes \nabla \phi$,
\begin{align}
    \nabla\cdot\pmb{\sigma}_\mathrm{K} &= \nabla \cdot(T\hat{\pmb{\sigma}}) = \hat{\pmb{\sigma}}\cdot \nabla T + \kappa_\phi T\cdot \left(\nabla\cdot\left(\Gamma(\nabla\phi) \frac{\partial\Gamma(\nabla\phi)}{\partial\nabla\phi}\right) \nabla \phi + \left(\Gamma(\nabla\phi) \frac{\partial\Gamma(\nabla\phi)}{\partial\nabla\phi} \cdot\nabla\right)\nabla\phi\right), \\
    &=\hat{\pmb{\sigma}}\cdot \nabla T + \kappa_\phi T\cdot \left(\nabla\cdot\left(\Gamma(\nabla\phi) \frac{\partial\Gamma(\nabla\phi)}{\partial\nabla\phi}\right)  \nabla \phi + \frac{1}{2}\nabla\Gamma^2(\nabla\phi )\right). \label{ko1}
\end{align}
Moreover, defining
\begin{equation}
    \mu = - \kappa_\phi\nabla\cdot\left(\Gamma(\nabla\phi) \frac{\partial\Gamma(\nabla\phi)}{\partial\nabla\phi}\right) + \frac{1}{T}\frac{\partial f}{\partial \phi},
\end{equation}
\eqref{ko1} can be formulated as
\begin{align*}
  \nabla\cdot\pmb{\sigma}_\mathrm{K} = \hat{\pmb{\sigma}}\cdot \nabla T - T\mu\nabla\phi +  \frac{\kappa_\phi}{2}T\nabla\Gamma^2(\nabla\phi) + \frac{\partial f}{\partial \phi}\nabla\phi.
\end{align*}
Using the reverse chain rule, we obtain
\begin{align*}
  \nabla\cdot\pmb{\sigma}_\mathrm{K} = \hat{\pmb{\sigma}}\cdot \nabla T - T\mu\nabla\phi +  \frac{\kappa_\phi}{2}T\nabla\Gamma^2(\nabla\phi) + \nabla f(T,\phi) - \frac{\partial f}{\partial T}\nabla T.
\end{align*}
Next, we employ the definition of total entropy $S:=-\frac{\partial f}{\partial T}-\frac{\kappa_\phi}{2}\Gamma^2(\nabla\phi)$ and the total Helmholtz free energy $F:=f+ \frac{\kappa_\phi}{2}T\Gamma^2(\nabla\phi)$ to obtain
\begin{align*}
  \nabla\cdot\pmb{\sigma}_\mathrm{K} = \hat{\pmb{\sigma}}\cdot \nabla T - T\mu\nabla\phi +  S(T,\phi) \nabla T  + \nabla F(T,\phi). \label{ko2}
\end{align*}
Here, we note that in the case of isotropic interfacial energy, the computation is exactly the same. However, in this case, $\mu$ is defined as 
\begin{equation}
    \mu = - \kappa_\phi\Delta\phi + \frac{1}{T}\frac{\partial f}{\partial \phi}.
\end{equation}

\clearpage
\bibliography{reference}%

\end{document}